\documentclass[a4paper,11pt]{article}
\pdfoutput=1 

\usepackage{jinstpub}
\usepackage{lineno}
\usepackage{textgreek}
\usepackage[separate-uncertainty=true]{siunitx}
\usepackage{subcaption}
\usepackage{tikz}

\notoc
\title{\boldmath Characterization measurements of the TRISTAN multi-pixel silicon drift detector}

\author[a,b,1]{K. Urban,\note{Corresponding author.}}
\author[c,d]{M.~Carminati,}
\author[e]{M.~Descher,}
\author[a,b]{F.~Edzards,}
\author[a]{D.~Fink,}
\author[c,d]{C.~Fiorini,}
\author[c,d]{M.~Gugiatti,}
\author[e]{D.~Hinz,}
\author[a,b]{T.~Houdy,}
\author[c,d]{P.~King,}
\author[f]{P.~Lechner,}
\author[a,b]{S.~Mertens,}
\author[a,b]{D.~Siegmann,}
\author[e]{M.~Steidl,}
\author[e]{J.~Wolf}

\affiliation[a]{Max-Planck-Institut für Physik, München, Germany}
\affiliation[b]{Physik-Department, Technische Universität München, Garching, Germany}
\affiliation[c]{DEIB, Politecnico di Milano, Milano, Italy}
\affiliation[d]{INFN, Sezione di Milano, Milano, Italy}
\affiliation[e]{Karlsruher Institut für Technologie, Eggenstein-Leopoldshafen, Germany}
\affiliation[f]{Halbleiterlabor der Max Planck Gesellschaft, München, Germany}

\emailAdd{korbinian.urban@tum.de}

\abstract{
Sterile neutrinos are a minimal extension of the Standard Model of Particle Physics. A laboratory-based approach to search for this particle is via tritium~\textbeta-decay, where a sterile neutrino would cause a kink-like spectral distortion. The Karlsruhe Tritium Neutrino~(KATRIN) experiment extended by a multi-pixel Silicon Drift Detector system has the potential to reach an unprecedented sensitivity to the keV-scale sterile neutrino in a lab-based experiment. The new detector system combines good spectroscopic performance with a high rate capability. In this work, we report about the characterization of charge-sharing between pixels and the commissioning of a 47-pixel prototype detector in a MAC-E filter. 
}


\proceeding{12th International Conference on Position Sensitive Detectors - PSD12\\
12-17 September, 2021, University of Birmingham}

\begin{document}
\maketitle

\section{Introduction}
Sterile neutrinos are a minimal extension of the Standard Model of particle physics, in which one or more additional neutrino mass eigenstates are introduced. The new mass eigenstate couples to the well-known active neutrino flavor eigenstates only via a small mixing. Depending on the mass of the new eigenstate~$m_{\text{s}}$, there are several motivations for the existence of sterile neutrinos. In case~$m_{\text{s}}$ is in the~\si{keV} range, sterile neutrinos are a viable dark matter candidate~\cite{boyarsky_19}.
\\\\
The kinematics of~\textbeta-decay can be used to probe the neutrino mass eigenstates in a laboratory experiment. The Karlsruhe Tritium Neutrino~(KATRIN) experiment makes use of this approach and determines the mass of the active neutrinos via a precision measurement of the endpoint region of the electron spectrum from tritium~\textbeta-decay~\cite{aker_19,aker_21}. A sterile mass eigenstate in the~keV-regime would lead to a kink-like signature at the energy~${E_0-m_{\text{s}}}$ in the spectrum, where~$E_0\approx \SI{18.6}{keV}$ is the endpoint of the electron spectrum~\cite{mertens_19}. To detect such a signature, the spectral energy range measured by KATRIN must be extended from the endpoint region to the entire spectrum. This requires a new detector system for \textbeta-spectroscopy, which is currently being developed within the TRISTAN project. The new detector system needs to have an electron rate capability of~$10^8\,$cps. At the same time, excellent spectroscopic properties, like energy resolution and linearity, are required. To achieve these goals, we use the silicon drift detector~(SDD) technology which allows an energy resolution of~\SI{300}{eV}~FWHM at~\SI{20}{keV} at high rates of~\SI{100}{kcps} per pixel. The detector system will be implemented as a multi-pixel focal plane array covering an area of about~${20\times 20~\si{cm^2}}$ with 3486~pixels of~\SI{7}{mm^2} each. The development of the detector system follows a staged approach. Starting with 7-pixel detector prototypes with a simple mechanical design, the detector chip was scaled to a more complex module consisting of 47~pixels, which has already been tested successfully. The final focal plane array will consist of~9~(phase-1) and 21~(phase-2) detector modules of 166-pixels each.

%
\begin{figure}[b]
\centering 
\input{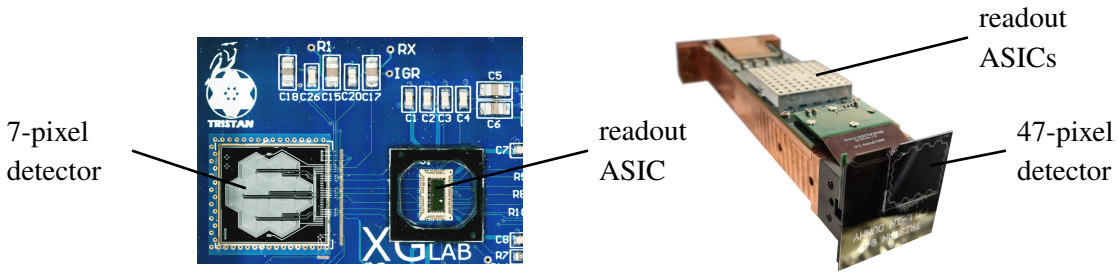}
\vspace*{-7mm}
\caption{\label{fig:det_pictures} Pictures of the 7-pixel and 47-pixel TRISTAN SDD detectors. The 47-pixel detector chip is mounted frontal on a copper block, perpendicular to the read-out electronics.}
\end{figure}
\section{Detector and readout scheme}
The TRISTAN detector follows the general design idea of SDDs used for X-ray spectroscopy~\cite{lechner_01}. 
Each detector chip is made of up to 166~seamless hexagonal pixels with \SI{3}{mm}~diameter each. The anode of every pixel is read out by a charge sensitive amplifier with a JFET transistor integrated into the anode structure of the chip, followed by a low-noise application-specific integrated circuit~(ASIC) specifically developed for the TRISTAN project~\cite{trigilio_18}. The integrated JFET allows one to place the ASIC chip at several~\si{cm} distance to the detector chip, while keeping the total anode capacity at only~\SI{180}{fF}. This provides an excellent signal to noise ratio
and enables the operation at room temperature. Cooling the detector up to \SI{-50}{\celsius} can further improve its performance. 
\textbf{Fig.}~\ref{fig:det_pictures} shows two detector setups: A 7-pixel detector in a planar electronics board configuration was used for X-ray charge sharing measurements.
For electron characterization measurements, a mechanically more complex 47-pixel detector, following the design of a final TRISTAN detector module, was used. In this design, all parts of the module are arranged behind the detector chip~(area of $38 \times \SI{40}{mm^2}$), so that 21~modules can be lined up with minimal distance to build a focal plane array. The 47-pixel detector module is an intermediate step toward the 166-pixel module.

\section{Characterization of charge sharing}
\begin{figure}[b]
\centering
\begin{minipage}{.6\textwidth}
  \centering
  \includegraphics[height=4.5cm]{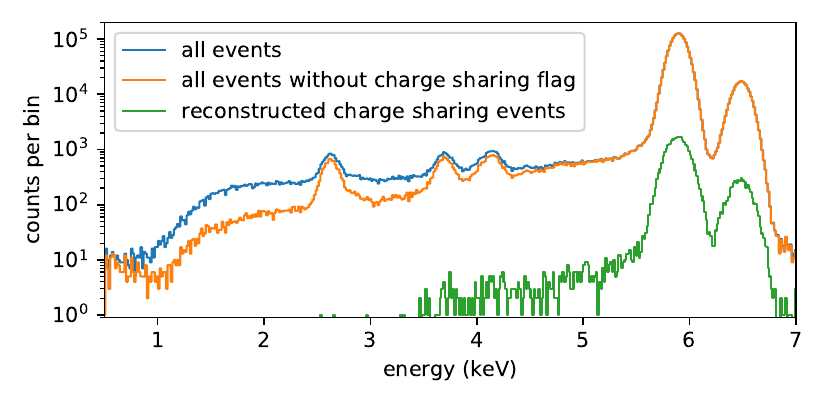}
  \captionof{figure}{Recorded spectrum of the central pixel in the presence of an~$^{55}\rm{Fe}$ source. The charge sharing-tagged events contribute to the low-energy tail and can be reconstructed.}
  \label{fig:cs_multiplicity_spectrum}
\end{minipage}
\hspace*{2mm}
\begin{minipage}{.37\textwidth}
  \centering
  \includegraphics[height=4.5cm]{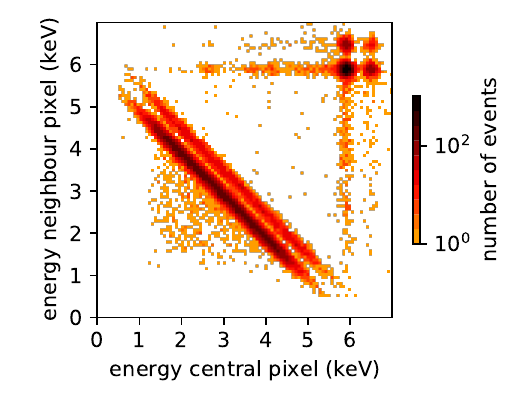}
  \captionof{figure}{Energy distribution of charge sharing-flagged events \\~}
  \label{fig:cs_2d_map}
\end{minipage}
\end{figure}
Due to the seamless and monolithic arrangement of the pixels, the charge cloud of an event at the pixel edge will be split between two adjacent pixels. If each pixel is read out independently, this adds a low-energy tail to the detector response, as part of the full charge cloud of an event is collected by the neighbouring pixel.
To quantify this effect, events in a calibration measurement with an~$^{55}\rm{Fe}$ X-ray source were acquired with a 7-pixel TRISTAN detector at room temperature. As the central pixel is fully surrounded by six neighboring pixels, the time coincidence with the neighbours can be used to identify charge sharing events. A coincidence window with a duration of~\SI{200}{ns} was chosen. \textbf{Fig.}~\ref{fig:cs_multiplicity_spectrum} shows the recorded $^{55}\rm{Fe}$~energy spectrum of the central pixel. By removing charge sharing-tagged events, the low-energy tail in the energy spectrum is reduced significantly. The energy distribution of both events causing the time coincidence is shown in~\textbf{Fig.}~\ref{fig:cs_2d_map}.
The energies detected with the central pixel and the neighboring pixels add up to the full X-ray energy. By adding both energies, the $^{55}\rm{Fe}$~spectrum can be reconstructed. In this particular measurement, about~\SI{2.1}{\percent} of the events in the central pixel show charge sharing. Using the pixel geometry and the observed energy threshold of~\SIrange{1.3}{1.5}{\keV}, this can be related to a Gaussian charge cloud with~$\sigma=\SI{11\pm1}{\micro m}$.

\section{Module characterization with electrons}
\begin{figure}[b]
\centering 
\input{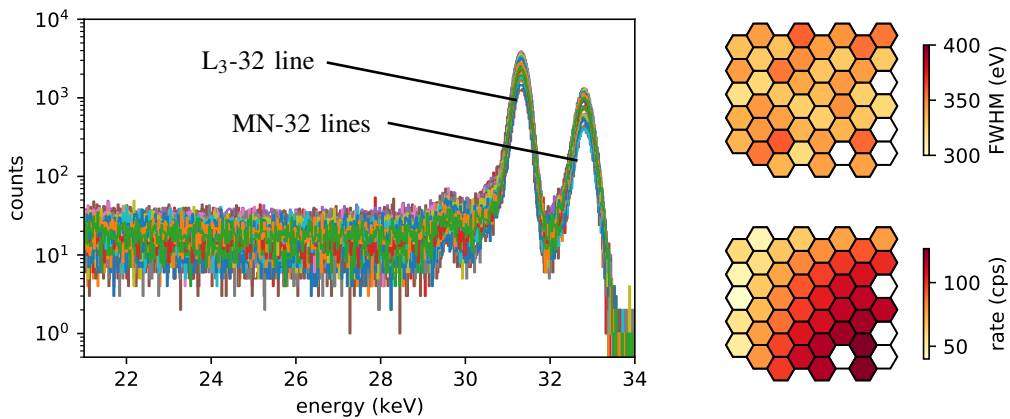}
\caption{Energy spectra of 43 out of the 47~pixels which were read-out at the monitor spectrometer setup. The two peaks correspond to L$_3$-32 and MN-32 conversion electron lines from~$^{\rm{83m}}\rm{Kr}$. The pixel maps show the energy resolution~(FWHM) and the total rate of the L$_3$-32~line.}
\label{fig:kr_spectra}
\end{figure}

The 47-pixel TRISTAN detector module was tested at the monitor spectrometer beamline of the KATRIN experiment. 
Similar to the main beamline of KATRIN, an electron source is combined with a kinetic energy filter~(MAC-E filter) followed by the detector. The electron source is a condensed, radioactive $^{\rm{83m}}\rm{Kr}$~source providing several monoenergetic electron lines from internal conversion~\cite{venos_18}. The source is set to a negative potential of~\SI{1000}{V}, which adds \SI{1}{\keV} to the kinetic energy of the electrons. In the measurement, the MAC-E filter was set to reject all electrons below the L$_3$-32~line at~\SI{31.47}{keV}, yielding a monoenergetic line from L$_3$-32~conversion at~\SI{31.47}{keV} and a superposition of L-32 and MN-32 conversion lines at around~\SI{33}{keV}. For data acquisition we used the Kerberos system, a 48-channel analog pulse processing and data acquisition platform \cite{king_21}.
The acquired spectra are shown in~\textbf{Fig.}~\ref{fig:kr_spectra}.
The detector was cooled to around \SI{-50}{\celsius}. All pixels show a similar performance and detector response. 
The rate on the detector varied between~\SIrange{40}{127}{cps} due to the magnetic field configuration at the beamline. The L$_3$-32~line is used to determine the energy resolution which is homogeneously distributed between~\SIrange{320}{360}{eV} FWHM over the entire detector chip.
\section{Conclusion}
Measurements of a 7-pixel TRISTAN detector prototype were analyzed regarding the charge sharing effect of neighboring pixels. Time coincidence could be used to obtain the energy distribution of charge sharing events. Furthermore, a mechanically more complex 47-pixel TRISTAN module showed the excellent performance of the SDD technology for high resolution \textbeta-spectroscopy in a realistic environment. The obtained energy resolution is better than~\SI{360}{eV} FWHM for~\SI{30}{keV} electrons on all read-out pixels. This new detector system, planned to be scaled to a~${>\num{3000}}$~pixel focal plane array, will enable the search for keV-scale~sterile neutrinos at the KATRIN experiment.

\bibliography{literature}

\end{document}